# Title: Moiré enabled spin pumping and preservation in $MoSe_2/WS_2$ heterobilayers

**Authors:** Leo Yu[1,2*], Kateryna Pistunova[1,2], Sudipta Kundu[3†], Jenny Hu[1], Kenji Watanabe[4], Takashi Taniguchi[5], Felipe H. da Jornada[2,3] and Tony F. Heinz[1,2]

**Affiliations:**

[1]E. L. Ginzton Laboratory, Stanford University, Stanford, CA 94305, USA.

[2]Stanford PULSE Institute, SLAC National Accelerator Laboratory, Menlo Park, CA 94025.

[3]Department of Materials Science and Engineering, Stanford University, Stanford, CA, 94305, USA.

[4]Research Center for Electronic and Optical Materials, National Institute for Materials Science, 1-1 Namiki, Tsukuba 305-0044, Japan.

[5]Research Center for Materials Nanoarchitectonics, National Institute for Materials Science, 1-1 Namiki, Tsukuba 305-0044, Japan.

†Current address: Department of Materials Science and Metallurgy, University of Cambridge, Cambridge, UK.

*Corresponding author. leoyu@stanford.edu

**Abstract:** The spin degree of freedom is a fundamental quantum mechanical attribute with implications spanning from magnetism to quantum computing. Consequently, the relaxation of spin states for extended, Bloch electrons in solids has been studied for decades as it defines many of their properties and applications. We show that moiré patterns in layered materials can extend spin relaxation times by two orders of magnitude to 1 millisecond and beyond. This is achieved by suppressing spin mixing for electrons in 2D semiconductor heterostructures, particularly in the $MoSe_2/WS_2$ system, as we elucidate both experimentally and theoretically. The extended longitudinal lifetime facilitates spin alignment over 50% using only nanowatt levels of optical power. Our findings highlight the potential of moiré engineering for future quantum sensing and information processing.

**Main Text:**

The relaxation of spin states for mobile Bloch electrons remains a fundamental challenge in solid-state physics (*1*, *2*), motivating renewed investigation with the advent of each novel material platform. Early research found spin relaxation times (SRTs) in metals limited by the Elliott-Yafet (EY) mechanism (*1*), which links spin relaxation to carrier momentum scattering. Subsequent decades saw advancements in semiconductors, where momentum scattering was reduced because of lower density of states and carrier concentration. By the 2000s, SRTs of 100 ns were achieved in GaAs (*3*) and Si (*4*), positioning "spintronics (*5*)" as a promising field and spurring further initiatives to mitigate spin relaxation. The localization of charge carriers quenches relaxation-inducing electron motions, allowing spins in quantum dots and charged impurities to exhibit extended coherence time required for qubit operations (*6*). For mobile electrons, a new strategy involves locking the spin to a quantum number that does not simultaneously flip, which can be naturally represented by the valley degree of freedom of transition metal dichalcogenides (TMDCs) semiconductors (*7*, *8*). While initial efforts using excitons in TMDC monolayers (MLs) were impeded by additional exchange-induced relaxation mechanisms (*9*), subsequent demonstrations based on resident carriers achieve spin/valley lifetimes of nanoseconds for electrons in $MoS_2$ (*10*) and $MoSe_2$ (*11*, *12*), and microseconds for holes in $WSe_2$ (*13*), eventually reaching 10 μs (*14*) for both carrier types in $WSe_2$.

Moiré superlattices in van der Waals (vdW) heterostructures (*15*) significantly alter electronic structures through the combined effect of zone folding, band hybridization, and atomic reconstruction (*16*, *17*). The resulting moiré flat bands impact charge physics profoundly, as seen in strong electronic correlations (*18*) and moiré excitons (*19*). Given the advantageous spin-valley locking property of TMDC MLs, an intriguing opportunity arises to further suppress spin relaxation by combining this locking with carrier confinement in moiré superlattices, as we demonstrate in this study. We also augment pump-probe spectroscopy with time-correlated single-photon counting, exploring a broad range of experimental parameters to achieve a fundamental mechanistic understanding. Our theoretical analysis integrates a conventional $\mathbf{k} \cdot \mathbf{p}$ approach with density-functional theory to consider complex moiré band structures not examined previously.

Our device comprises angle-aligned $MoSe_2/WS_2$ heterobilayers (HBLs) and additional adjacent $MoSe_2$ MLs. The HBL exhibits a type-I band alignment between the two constituent materials as established in prior studies (*20*, *21*). Both the conduction band minimum (CBM) and valance band maximum (VBM) of the HBL reside in the lower-gap $MoSe_2$ layer, determining the lowest-energy optical transitions to be intralayer excitons instead of interlayer ones. Comparative optical studies can thus be done between the HBL and the ML. On the other hand, the $WS_2$ layer, with its 4%-smaller lattice constant, introduces a moiré superlattice with a periodicity of 8 nm. The resulting lattice reconstruction (*22*) (Fig. 1a) leads to confined electronic wavefunctions as will be shown later. The HBL and the ML are encapsulated in insulating hexagonal boron nitride (h-BN) layers and equipped with dual gates for doping control and electric tuning, as depicted in Fig. 1b. We focus on two HBL samples in this study with twist angles close to $60^{o}$ or $55^{o}$, respectively, corresponding to an *H*-type stacking order.

The doping-dependent reflection contrast (RC) spectra from the HBL reveal complex exciton structures that are absent in the ML. The ML displays relatively simple spectral characteristics, with a neutral exciton peak in the intrinsic region and a trion peak when a positive gate voltage is applied to introduce additional electrons (Fig. 1c). In contrast, the HBL exhibits various spectral features at equally spaced positive voltages, indicated by dashed lines (Fig. 1d), which correspond to integer fillings of moiré superlattices ranging from 0 to 3 electrons per moiré site, consistent with recent studies (*20*). At the particular filling factor ν = 0, we observe the fundamental (1.60 eV) exciton $X_1$, as well as an additional peak $X_2$ (1.64 eV) consistent with previous reports (*23*) on moiré excitons. At the filling factor ν = 1, the new feature $X_T$ (at 1.57 eV) is identified as an intervalley trion species (supplementary text section 1). These rich features will facilitate pump-probe spectroscopy. Furthermore, a comparison of the trion peak between the ML and HBL shows similarities in their binding energies and oscillator strengths, suggesting that the charge carriers in both systems are mobile rather than localized, consistent with prior electrical transport studies (*24*).

**Polarization of valley electrons via magnetic field and optical pumping**

TMDC MLs feature two nonequivalent electronic valleys (*8*, *9*) ($\mathbf{K}$ and $\mathbf{K}'$) that are equally populated by resident electrons in equilibrium. The control of valley polarization, or the population imbalance, has been achieved using an external magnetic field (*8*, *9*), which we show in the HBL for subsequent calibration. At an applied field of 2 T, the Zeeman splitting between the two valleys exceeds the thermal energy at our base measurement temperature (~ 1.8 K), causing the resident electrons to predominantly populate the lower-energy valley (supplementary text section 1). The polarization of valley electrons can be optically observed thanks to the valley circular dichroism inherent in TMDC MLs and the trion nature of the $X_T$ peak. Specifically, we measure spectrally resolved light ellipticity in reflective helicity contrasting (RHC) spectra of circularly polarized whitelight, defined as $\left(I_{\sigma_+}(\omega) - I_{\sigma_-}(\omega)\right) / \left(I_{\sigma_+}(\omega) + I_{\sigma_-}(\omega)\right)$. As shown in Fig. 1e, the RHC spectrum displays significant contrast at the $X_T$ peak and reverses its sign with the orientation of the magnetic field. The peak value sets an RHC reference for complete valley polarization. We further calibrate the measurement of valley polarization in light ellipticity in supplementary text section 1.

In addition to magnetic control, we demonstrate that substantial valley polarization can also be achieved in the HBL via optical pumping. Fig. 1f presents a representative RHC spectrum at ν = 0.8, with the magnetic field turned off but a continuous-wave (c.w.) pump light tuned to the $X_2$ peak and set above the saturation power (20 nW). The RHC spectrum changes sign with the helicity of the pump light and generally resembles that induced by the magnetic field in Fig. 1e. By comparing the peak RHC induced by the pump light and by the magnetic field, we determine the fraction of valley electrons polarized via optical spin pumping. We calculate the pump-induced valley polarization at a given doping level from the doping-dependent RHC spectra in Fig. 1i, j. A polarization over 50% can be achieved at ν = 0.4-0.7, as shown in Fig. 1l. This high degree of valley polarization, while reported (*25*) in monolayer $WSe_2$ and $WS_2$ whose CBM and VBM are spin-anti-aligned, has not been reported in $MoS_2$ (*26*) nor in $MoSe_2$ (*12*) whose CBM and VBM are spin-aligned.

Similar measurements of valley polarization were conducted in the ML for comparison, as shown in Fig. 1g, h. The magnetic-field-induced RHC in the ML is comparable to that in the HBL. But the pump-induced RHC in the ML is negligible within experimental uncertainty, even with a pump power nearly four orders of magnitude higher than that used for the HBL. This lack of appreciable spin pumping even at high pump powers suggests a significant difference in spin relaxation times between the ML and the HBL.

**Dynamics of spin relaxation in heterobilayer and monolayer $MoSe_2$**

The substantial valley polarization achievable in the HBL suggests a slow spin relaxation dynamics, which we explore by measuring time-resolved light ellipticity through pump-probe spectroscopy (TRLE, Materials and Methods and supplementary text section 2). The use of time-correlated single-photon counting (Materials and Methods) further enables access to extended time delays, even with probe light at the single-photon level. We use a pulsed pump beam resonant with the $X_2$ peak and a c.w. probe beam resonant with the $X_T$ peak, initially activating them separately as test cases (Fig. 2a). When only the probe beam is on, the probe light intensity shows no time-dependent modulation on the detector. When only the pump beam is on, excitons are excited. Some excitons radiatively recombine and emit at the $X_T$ energy, producing a photoluminescence (PL) signal on the detector, albeit with an intensity significantly lower than that of the probe light and with a duration indistinguishable from the instrument response. Note the pump light at the $X_2$ energy is blocked from the detector by optical filters.

When both the pump and probe beams are on, a notable modulation of the probe light intensity is detected (Fig. 2a). This modulation gradually intensifies from the onset of the pump pulse (Fig. 2d) and persists for several hundred microseconds beyond the termination of the pump pulse. The intensity modulation can be set positive or negative by selecting the helicity of the pump beam. Both the temporal scale and the sign change of the intensity modulation provide insights into its potential causes. PL from trions and residual pump light can be excluded as potential causes, as they neither induce a negative intensity modulation nor persist for hundreds of microseconds. We therefore attribute the cause to the valley polarization of resident electrons. The gradual buildup of the valley/spin polarization further demonstrates the spin pumping effect. We note the TRLE signal can be shown to be proportional to the RHC in supplementary text section 1.

In contrast, we observe a significantly faster spin relaxation in the ML. Upon exciting the ML with pump light above its bandgap, the ellipticity signal decays over time with a time constant of 2.4 ns (Fig. 2b). The sign of the ellipticity signal follows the helicity of the pump light. These observations are consistent with previous reports (*11*, *12*) of spin relaxation times of 2-4 ns in ML $MoSe_2$, although they fall short of an 100-ns report (*27*) using optimal crystals from various suppliers. We note that the h-BN encapsulation of our device, not present in previous studies of $MoSe_2$ MLs, does not extend the spin relaxation time. The ineffectiveness of the encapsulation is consistent with reports on ML $WSe_2$ (*14*).

Our TRLE data do not readily facilitate a microscopic understanding of the optical pumping process. However, we reckon that the efficiency of angular momentum transfer from the pump photons to the resident electrons is at least 2% and likely much higher (supplementary text section 3), given that near unity valley polarization was achieved in the HBL with only nanowatts of pump power. The process likely involves exciton correlations and scattering processes in the presence of photogenerated minority carriers (*13*), but it is beyond the scope of this study. Regarding the differences between the HBL and the ML, the disparity in spin relaxation times by orders of magnitude supports the absence of appreciable valley polarization in the latter case (Fig. 1). Also, we note that the multitude of exciton peaks in the HBL permits resonant pumping via the $X_2$ and the $X_1$ peak (supplementary text section 3), whereas the pumping for the ML was done non-resonantly in our study and in previous measurements (*11*, *12*).

**Relaxation-driving carrier scattering in moiré heterobilayers**

The notably prolonged spin relaxation time of moiré electrons in $MoSe_2$ necessitates an investigation of the relaxation mechanism, beginning with identifying various types of carrier scattering from the relaxation rate dependence on various parameters including temperature, doping, and optical power. First, we examine the temperature dependence of the spin relaxation rate in the low-doping (ν = 0.2) regime (Fig. 3a, b), where a pronounced decrease is observed around 10 K, below which the rate essentially remains constant. This temperature dependence is well-captured by an Arrhenius fit, yielding an activation energy of 17 meV. This energy closely aligns with that of zone-edge phonons, leading us to attribute the carrier scattering type to an electron-phonon one. We note that phonon-mediated intravalley scattering, although widely considered (*28*, *29*), is not observed in our temperature dependence data nor in measurements of other monolayer systems (*14*) (supplementary text section 5).

At the base temperature (~ 1.8 K), we then assess the doping dependence of the spin relaxation rate (Fig. 3c, d) across the range of ν = 0.1-0.9, beyond which the effect of strong electronic correlation prevails and is left to future studies. Two regimes of doping dependence are discernible around ν ~ 0.3 (Fig. 3d). Above ν ~ 0.3, the rate monotonically increases with the filling factor. As phonons are frozen out, we expect the relaxation to be driven by impurity-mediated intervalley scattering (*7*, *30*). The observed rate increase with carrier density indeed agrees with theoretical predictions (*31*) of such carrier scattering. It is important to note that exchange intervalley electron-electron scattering is prohibited in TMDC MLs in the regime where spin splitting is greater than Fermi energy (*32*). Also, the matrix elements of spin-orbit interaction in TMDC MLs are independent of carrier momentum, in contrast to that in the Kane model of III-V semiconductors, as we will see later. At lower carrier density (ν = 0.3 or below), a further decrease in the rate is anticipated.

While we expect the spin relaxation rate to be eventually limited by impurity scattering at the lowest controllable carrier density, we find the limit obscured by additional photo-induced relaxation mechanisms in optical-power-dependent studies. We find the relaxation rate independent of the power of the pump light, as its duration is only 1 μs. However, the rate exhibits a linear dependence on the probe power, ranging from 1.5 to 50 nW (Fig. 3e, f) at two representative filling factors of ν = 0.2 and 0.8 (not shown). We attribute the origin of the photo-induced relaxation to residual optical pumping from the probe laser, which, although mitigated by using the $X_T$ peak, remains non-negligible. We note that photo-induced heating or doping effects are negligible in our device at the optical power levels studied (supplementary text section 1 and supplementary text section 4). We also note that the observed rate dependence on temperature, doping, and optical power is qualitatively reproduced in the additional, 55º-aligned sample (supplementary text section 4). We discuss other spin relaxation mechanisms and their incompatibility with our experimental findings in supplementary text section 5.

Despite the presence of extrinsic relaxation mechanisms, we measure a spin relaxation time of 820 μs at a probe power of 1.5 nW (Fig. 3e), and we can infer a relaxation time approaching 2 ms by extrapolating the fit to the power-dependent relaxation rate (Fig. 3f). The measured spin relaxation time exceeds the previous record (40 μs (*33*)) in TMDC semiconductors in unaligned $WSe_2/MoS_2$ HBLs by more than tenfold, is about a hundredfold longer than that in $WSe_2$ monolayers (10 μs (*14*)) at comparable carrier densities (< $5*10^{11}$ $cm^{-2}$), and is about five orders of magnitude longer than that in $MoSe_2$ monolayers (2-4 ns (*11*, *12*)). The extended spin relaxation times are also observed in a less aligned, 55º- sample: ~ 100 μs in measurements and 200 μs in extrapolation (supplementary text section 4). The trend of shorter relaxation times in the less aligned HBLs and in the $MoSe_2$ monolayers suggests the beneficial role of moiré confinement in suppressing spin relaxation.

**Theory of spin relaxation in TMDC MLs**

To elucidate the origin of the suppressed spin relaxation in moiré HBLs, we first consider the simpler problem of spin relaxation in TMDC monolayers. It is well-established that four main mechanisms govern spin relaxation for mobile electrons in a solid (*1*): the Elliott-Yafet (EY) mechanism, the D'yakonov-Perel' (DP) mechanism, the Bir-Aronov-Pikus (BAP) mechanism, and the hyperfine-interaction mechanism. We posit that the latter three mechanisms are unimportant in our system and the EY mechanism dominates. Briefly, the BAP mechanism, mediated by exchange interaction, is irrelevant in the electron-doped samples under study (*34*). The nuclear hyperfine interaction is generally weak and further diminished by carrier motional narrowing (*34*).

Also constraining spin relaxation in TMDC MLs is the symmetry of the band edge states at $\mathbf{K}$ and $\mathbf{K}'$ point, represented by the $C_{3h}$ point group, especially through the mirror-reflection operator $\sigma_h$ defined with respect to the horizontal plane spanning the metal atoms (*28*, *34*). A consequence of having $\sigma_h$ in the point group is the requirement for spin mixing to occur solely between an even and an odd band, as the (out-of-plane)-spin-flip operators $L_\pm S_\mp$ are odd against $\sigma_h$. In TMDC MLs, the conduction ($c$) and valance ($v$) bands are even but the $c+1$ and $v-1$ bands are odd (*35*). Orbital quantum number further dictates that the CBM in the $c\downarrow$ band only spin-mixes with the $v-1\uparrow$ band (*34*). Another related consequence of having $\sigma_h$ is the protection of out-of-plane spin states (*29*, *34*), unless the mirror symmetry is broken explicitly by a perturbative Hamiltonian. Given that the intrinsic spin-orbit field of TMDC MLs points predominantly out-of-plane, the suppression of the in-plane field renders the precession-driven DP mechanism ineffective. Further examination of alternative relaxation channels, summarized in supplementary text section 5, shows their incompatibility with experimental findings. These alternatives include the DP mechanism induced by the Rashba effect (*29*, *34*, *36*), long-wavelength acoustic phonons (*28*, *29*, *36*, *37*), magnetic impurities, and conduction-band hybridization in the HBLs.

The experimental observations and theoretical considerations enable us to ascertain the mechanism of spin relaxation in TMDC MLs at low temperature: the EY mechanism driven by impurity scattering. In this framework, the intervalley matrix element is proportional to the mixing between opposite spin states, which Habe and Koshino (*31*) have further suggested scales linearly with the state separation from $\mathbf{K}$ point. We derive the intervalley matrix element formally using a seven-band $\mathbf{k}\cdot\mathbf{p}$ model established by Kormányos et al. (*35*). Our expression not only incorporates concrete band parameters but will also be generalized for moiré HBLs later:

$$\langle \psi_{c\uparrow\mathbf{K}'+\mathbf{q}'}|\, V_{\text{imp}}\, |\psi_{c\downarrow\mathbf{K}+\mathbf{q}}\rangle = \frac{\Delta_{c,v-1}}{\epsilon_c-\epsilon_{v-1}}\frac{\gamma_3(q_-+q'_-)}{\epsilon_c-\epsilon_v}M^{v-1,v}_{\mathbf{K}'\mathbf{K}} = a_{v-1\uparrow,\,\text{ML}}a_{v\downarrow,\,\text{ML}}M^{v-1,v}_{\mathbf{K}'\mathbf{K}}$$

Here, $\mathbf{q}$ and $\mathbf{q}'$ are the carrier momentum of the initial and final states away from the band extremum $\mathbf{K}$ and $\mathbf{K}'$. $V_{\text{imp}}(\mathbf{r})$ is an impurity Hamiltonian that conserves spin states. $\epsilon_b$ is the energy of the band $b$ of pristine TMDC MLs. $\gamma_3$ and $\Delta_{c,v-1}$ parametrizes, respectively, the $\mathbf{k}\cdot\mathbf{p}$ and the spin-orbit interaction between the $c$ and $v$ band and between the $c$ and $v-1$ band. $M^{v-1,v}_{\mathbf{K}'\mathbf{K}}$ is the impurity scattering matrix element between the band edge states. $q_-=q_x-iq_y$ is the symmetrized momentum invariant under the $C_{3h}$ group. In addition to the band parameters, the matrix element can also be expressed in terms of the projection amplitudes $a_{v-1\uparrow,\,\text{ML}}$ and $a_{v\downarrow,\,\text{ML}}$ onto the respective bands. Furthermore, our expression indicates a linear dependence of spin relaxation rate on carrier density $n$, since the Fermi wavevector $q_{\text{F}}=\sqrt{2\pi n}$. The trend agrees with previous theoretical predictions (*31*) and our measurements (Fig. 3), but diverges from the scaling in the Kane model (*1*). We include our derivation, the justification for omitting remote bands, an expression of the spin relaxation rate and its comparison with the Kane model result, along with more symmetry considerations including types of point defects, in supplementary text section 6.

**Spin relaxation in $MoSe_2/WS_2$ heterobilayers with strain-induced moiré potential**

As the intervalley matrix element is decomposable into band projection amplitudes for TMDC MLs, we see the matrix element for HBLs can be similarly calculated, contingent upon its $\mathbf{k}\cdot\mathbf{p}$ expansion near the CBM. We employ density-functional theory (DFT) to determine the moiré band structure of the HBL (*22*), identifying the CBM location at $\mathbf{K}_{\mathrm{M}}$ (Fig. 4a), where the subscript $_{\mathrm{M}}$ denotes the wavevector in the moiré Brillouin zone (MBZ) as opposed to the unit-cell Brillouin zone (UBZ). The $\mathbf{k}\cdot\mathbf{p}$ expansion for the HBL band structure has not been pursued due to the significantly larger number of bands within the same energy range (from $v-1$ to $c$ band) in the HBL (~5000) compared to the ML (3*2). Upon analyzing the wavefunction of each band, it becomes apparent that wavefunctions characterized by the same wavevector (say, $\mathbf{K}_{\mathrm{M}}$) in the MBZ generally carry disparate wavevectors in the UBZ. This disparity in inter-unit-cell phase evolution results in negligible overlap among most wavefunctions. As the CBM wavefunction exhibits a momentum distribution peaking at $\mathbf{K}$ in the UBZ (Fig. 4f), we identify about 60 wavefunctions that similarly peak at $\mathbf{K}$ or in its nearest neighborhood, which will be referred to as the KNN wavefunctions. We detail the Fourier analysis of DFT-calculated wavefunctions in supplementary text section 7.

Projecting a wavefunction near the CBM onto the KNN wavefunctions, we find the bases that carry the largest amplitudes can be labeled as $\{v\downarrow, v-1\uparrow\}\otimes\{1,4,7\}$, where the dominant ML-band composition is supplemented with a mini-band index distinguishing the spatial profile of an envelope function. The projection bases and their amplitudes together constitute the $\mathbf{k}\cdot\mathbf{p}$ expansion. Fig. 4b depicts the amplitudes onto the $v\downarrow$- mini-bands. For the ML, the amplitude grows linearly with carrier momentum, as expected from the expression $\gamma_3 q/(\epsilon_c-\epsilon_v)$. The extracted $\gamma_3$ is 3.42 eV·Å, reasonably close to the literature value of 2.53 eV·Å (*35*) given DFT implementation differences. For the HBL, while the momentum dependence remains similar, the amplitudes decrease appreciably from the ML value, a phenomenon that will be explained next.

As a result of lattice reconstruction (Fig. 1a) in a moiré supercell (*22*), strain varies slowly over nanometer length scales and becomes concentrated at specific high-symmetry stackings (Fig. 4d). Given that strain is known to shift the band-edge energy of TMDC semiconductors, to capture the resultant slow spatial variation of a wavefunction we introduce an envelope function $F_{bm_{\mathrm{M}}^b}(\mathbf{R})$, where $b$ follows the ML band index and $m_{\mathrm{M}}^b$ is the additional mini-band index associated with the band $b$. The label $\mathbf{R}$ indexes all the unit-cells within a moiré supercell and does not account for finer spatial coordinates in a unit-cell. Normalization requires $\sum_{\mathbf{R}}\left|F_{bm_{\mathrm{M}}^b}(\mathbf{R})\right|^2=1$. Extracting the envelopes from the wavefunctions, we see in Fig. 4 that $F_{c\downarrow1}(\mathbf{R})$ or $F_{v-1\uparrow1}(\mathbf{R})$ maximizes in regions of high tensile strain or high compressive strain, respectively. This contrasting confinement behavior is not coincidental but reflects the orbital and symmetry characteristics ($d_{z^2}$ versus $d_{xz,yz}$) of the underlying bands. We discuss the validity of our envelope function description in supplementary text section 7.

The reduction in projection amplitudes, and consequently in intervalley matrix elements, can be largely understood in terms of the overlap between the envelope functions of disparate mini-bands, i.e., $\left\langle F_{b'm_{\mathrm{M}}^{b'}}\middle|F_{bm_{\mathrm{M}}^b}\right\rangle \equiv \sum_{\mathbf{R}} F^*_{b'm_{\mathrm{M}}^{b'}}(\mathbf{R})F_{bm_{\mathrm{M}}^b}(\mathbf{R})$ . For example, in Fig. 4b, $a_{v\downarrow\mathbf{q}m_{\mathrm{M}}^v}\approx$

$a_{v\downarrow\mathbf{q},\,\mathrm{ML}}\langle F_{v\downarrow m_{\mathrm{M}}^{v}}|F_{c\downarrow 1_{\mathrm{M}}}\rangle$, which approximation is accurate with error below 3% and can be substantiated using perturbation theory (supplementary text section 7). The intervalley matrix element through the mini-band $v-1\uparrow m_{\mathrm{M}}^{v-1}$ and $v\downarrow m_{\mathrm{M}}^{v}$ is thus modified approximately by $\left\langle F_{v-1\uparrow m_{\mathrm{M}}^{v-1}}\middle|F_{c\downarrow 1_{\mathrm{M}}^{\mathrm{c}}}\right\rangle\cdot\langle F_{v\downarrow m_{\mathrm{M}}^{v}}|F_{c\downarrow 1_{\mathrm{M}}^{\mathrm{c}}}\rangle\cdot\left\langle F_{v-1\uparrow m_{\mathrm{M}}^{v-1}}\middle|F_{v\downarrow m_{\mathrm{M}}^{v}}\right\rangle$. The overall squared modulus of the matrix elements is obtained by summing over all combinations of mini-bands $(m_{\mathrm{M}}^{v-1}, m_{\mathrm{M}}^{v})$. This summation (supplementary text section 7) reveals a reduction in the squared modulus between the HBL and the ML by a factor of 7, consistent with our experimental observations of suppressed spin relaxation in the HBL. Achieving quantitative agreement would require more precise knowledge of the mirror-asymmetric $(v-1, c+1)$ bands (supplementary text section 7). On the one hand, the so-called all-electron DFT with less pseudization predicts up to twice the strain-induced shift for the $v-1$ band (*38*), although this approach is beyond the reach of current computational capabilities for large-scale moiré systems. At a more basic level, the energy of the asymmetric bands varies by ~100 meV in DFT literature (*38–40*) and could not be corroborated with experimental data. These essential issues in spin physics await collaborative efforts from the broad community.

## Moiré spintronics

Our findings indicate the potential to engineer moiré materials with optimal stacking configurations and energy level structures for spintronic applications. The advancement of coherent spin-valley operations (*10*) appears promising through the integration of recent theoretical proposals (*41–43*) with reduced scattering in moiré confinement. On the other hand, moiré superlattices exhibit rich, highly tunable correlated electronic phases (*16*, *17*), whose interplay with spin physics remains largely unexplored in conventional spintronic materials. Coupled with the unique exposedness and integrability of van der Waals materials, we anticipate new directions in quantum sensing using quantum materials (*44*, *45*).

## References and Notes

1. I. Žutić, J. Fabian, S. Das Sarma, Spintronics: Fundamentals and applications. *Rev. Mod. Phys.* **76**, 323–410 (2004).

2. J. Fabian, A. Matos-Abiague, C. Ertler, P. Stano, I. Žutić, Semiconductor spintronics. *Acta Phys. Slovaca* **57**, 565 (2007).

3. J. M. Kikkawa, D. D. Awschalom, Resonant Spin Amplification in n -Type GaAs. 4313–4316 (1998).

4. R. Jansen, Silicon spintronics. Nature Publishing Group (2012). https://doi.org/10.1038/nmat3293.

5. S. a Wolf, D. D. Awschalom, R. a Buhrman, J. M. Daughton, S. von Molnár, M. L. Roukes, a Y. Chtchelkanova, D. M. Treger, Spintronics: a spin-based electronics vision for the future. *Science* **294**, 1488–1495 (2001).

6. A. Chatterjee, P. Stevenson, S. De Franceschi, A. Morello, N. P. de Leon, F. Kuemmeth, Semiconductor qubits in practice. *Nat. Rev. Phys.* **3**, 157–177 (2021).

7. D. Xiao, G.-B. Liu, W. Feng, X. Xu, W. Yao, Coupled Spin and Valley Physics in Monolayers of MoS_{2} and Other Group-VI Dichalcogenides. *Phys. Rev. Lett.* **108**, 196802 (2012).

8. X. Xu, W. Yao, D. Xiao, T. F. Heinz, Spin and pseudospins in layered transition metal dichalcogenides. *Nat. Phys.* **10**, 343–350 (2014).

9. J. R. Schaibley, H. Yu, G. Clark, P. Rivera, J. S. Ross, K. L. Seyler, W. Yao, X. Xu, Valleytronics in 2D materials. *Nat. Rev. Mater.* **1**, 16055 (2016).

10. L. Yang, N. A. Sinitsyn, W. Chen, J. Yuan, J. Zhang, J. Lou, S. A. Crooker, Long-lived nanosecond spin relaxation and spin coherence of electrons in monolayer MoS 2 and WS 2. *Nat. Phys.* **11**, 830–834 (2015).

11. M. Schwemmer, P. Nagler, A. Hanninger, C. Schüller, T. Korn, Long-lived spin polarization in n-doped MoSe2 monolayers. *Appl. Phys. Lett.* **111**, 2–6 (2017).

12. R. R. Rojas-Lopez, F. Hendriks, C. H. van der Wal, P. S. S. Guimarães, M. H. D. Guimarães, Magnetic field control of light-induced spin accumulation in monolayer MoSe2. *2D Mater.* **10** (2023).

13. P. Dey, L. Yang, C. Robert, G. Wang, B. Urbaszek, X. Marie, S. A. Crooker, Gate-Controlled Spin-Valley Locking of Resident Carriers in WSe2 Monolayers. *Phys. Rev. Lett.* **119**, 1–5 (2017).

14. J. Li, M. Goryca, K. Yumigeta, H. Li, S. Tongay, S. A. Crooker, Valley relaxation of resident electrons and holes in a monolayer semiconductor: Dependence on carrier density and the role of substrate-induced disorder. *Phys. Rev. Mater.* **5**, 1–11 (2021).

15. A. K. Geim, I. V. Grigorieva, Van der Waals heterostructures. *Nature* **499**, 419–425 (2013).

16. E. Y. Andrei, D. K. Efetov, P. Jarillo-Herrero, A. H. MacDonald, K. F. Mak, T. Senthil, E. Tutuc, A. Yazdani, A. F. Young, The marvels of moiré materials. *Nat. Rev. Mater.* **6**, 201–206 (2021).

17. K. F. Mak, J. Shan, Semiconductor moiré materials. *Nat. Nanotechnol.* **17**, 686–695 (2022).

18. L. Balents, C. R. Dean, D. K. Efetov, A. F. Young, Superconductivity and strong correlations in moiré flat bands. *Nat. Phys.* **16**, 725–733 (2020).

19. D. Huang, J. Choi, C. K. Shih, X. Li, Excitons in semiconductor moiré superlattices. *Nat. Nanotechnol.* **17**, 227–238 (2022).

20. Y. Tang, J. Gu, S. Liu, K. Watanabe, T. Taniguchi, J. C. Hone, K. F. Mak, J. Shan, Dielectric catastrophe at the Wigner-Mott transition in a moiré superlattice. *Nat. Commun.* **13**, 1–7 (2022).

21. J. Kistner-Morris, A. Shi, E. Liu, T. Arp, F. Farahmand, T. Taniguchi, K. Watanabe, V. Aji, C. H. Lui, N. Gabor, Electric-field tunable Type-I to Type-II band alignment transition in MoSe2/WS2 heterobilayers. *Nat. Commun.* **15**, 4075 (2024).

22. M. H. Naik, E. C. Regan, Z. Zhang, Y. H. Chan, Z. Li, D. Wang, Y. Yoon, C. S. Ong, W. Zhao, S. Zhao, M. I. B. Utama, B. Gao, X. Wei, M. Sayyad, K. Yumigeta, K. Watanabe, T. Taniguchi, S. Tongay, F. H. da Jornada, F. Wang, S. G. Louie, Intralayer charge-transfer moiré excitons in van der Waals superlattices. *Nature* **609**, 52–57 (2022).

23. E. M. Alexeev, D. A. Ruiz-Tijerina, M. Danovich, M. J. Hamer, D. J. Terry, P. K. Nayak, S. Ahn, S. Pak, J. Lee, J. I. Sohn, M. R. Molas, M. Koperski, K. Watanabe, T. Taniguchi, K. S. Novoselov, R. V. Gorbachev, H. S. Shin, V. I. Fal'ko, A. I. Tartakovskii, Resonantly hybridized excitons in moiré superlattices in van der Waals heterostructures. *Nature* **567**, 81–86 (2019).

24. T. Li, J. Zhu, Y. Tang, K. Watanabe, T. Taniguchi, V. Elser, J. Shan, K. F. Mak, Charge-order-enhanced capacitance in semiconductor moiré superlattices. *Nat. Nanotechnol.* **16**, 1068–1072 (2021).

25. C. Robert, S. Park, F. Cadiz, L. Lombez, L. Ren, H. Tornatzky, A. Rowe, D. Paget, F. Sirotti, M. Yang, D. Van Tuan, T. Taniguchi, B. Urbaszek, K. Watanabe, T. Amand, H. Dery, X. Marie, Spin/valley pumping of resident electrons in WSe2 and WS2 monolayers. *Nat. Commun.* **12**, 5455 (2021).

26. S. Park, S. Arscott, T. Taniguchi, K. Watanabe, F. Sirotti, F. Cadiz, Efficient valley polarization of charged excitons and resident carriers in Molybdenum disulfide monolayers by optical pumping. *Commun. Phys.* **5**, 1–8 (2022).

27. M. Ersfeld, F. Volmer, P. M. M. C. De Melo, R. De Winter, M. Heithoff, Z. Zanolli, C. Stampfer, M. J. Verstraete, B. Beschoten, Spin States Protected from Intrinsic Electron-Phonon Coupling Reaching 100 ns Lifetime at Room Temperature in MoSe2. *Nano Lett.* **19**, 4083–4090 (2019).

28. Y. Song, H. Dery, Transport theory of monolayer transition-metal dichalcogenides through symmetry. *Phys. Rev. Lett.* **111**, 1–5 (2013).

29. H. Ochoa, F. Guinea, V. I. Fal'Ko, Spin memory and spin-lattice relaxation in two-dimensional hexagonal crystals. *Phys. Rev. B - Condens. Matter Mater. Phys.* **88** (2013).

30. K. Kaasbjerg, J. H. J. Martiny, T. Low, A.-P. Jauho, Symmetry-forbidden intervalley scattering by atomic defects in monolayer transition-metal dichalcogenides. *Phys. Rev. B* **96**, 241411 (2017).

31. T. Habe, M. Koshino, Spin relaxation in a hole-doped transition-metal dichalcogenide monolayer and bilayer with crystal defects. *Phys. Rev. B* **93**, 1–5 (2016).

32. D. Miserev, J. Klinovaja, D. Loss, Exchange intervalley scattering and magnetic phase diagram of transition metal dichalcogenide monolayers. *Phys. Rev. B* **100**, 1–9 (2019).

33. J. Kim, C. Jin, B. Chen, H. Cai, T. Zhao, P. Lee, S. Kahn, K. Watanabe, T. Taniguchi, S. Tongay, M. F. Crommie, F. Wang, Observation of ultralong valley lifetime in WSe2/MoS2 heterostructures. *Sci. Adv.* **3**, 1–7 (2017).

34. H. Ochoa, R. Roldán, Spin-orbit-mediated spin relaxation in monolayer MoS2. *Phys. Rev. B - Condens. Matter Mater. Phys.* **87**, 1–8 (2013).

35. A. Kormányos, G. Burkard, M. Gmitra, J. Fabian, V. Zólyomi, N. D. Drummond, V. Fal'ko, K · P Theory for Two-Dimensional Transition Metal Dichalcogenide Semiconductors . *2D Mater.* **2**, 022001 (2015).

36. A. J. Pearce, G. Burkard, Electron spin relaxation in a transition-metal dichalcogenide quantum dot. *2D Mater.* **4** (2017).

37. J. Xu, A. Habib, S. Kumar, F. Wu, R. Sundararaman, Y. Ping, Spin-phonon relaxation from a universal ab initio density-matrix approach. *Nat. Commun.* **11** (2020).

38. S. H. Rhim, Y. S. Kim, A. J. Freeman, Strain-induced giant second-harmonic generation in monolayered 2 H -MoX2 (X = S, Se, Te). *Appl. Phys. Lett.* **107** (2015).

39. C. Chang, X. Fan, Orbital analysis of electronic structure and phonon dispersion in MoS 2 , MoSe 2 , WS 2 , and WSe 2 monolayers under strain. **195420**, 1–9 (2013).

40. S. Puri, S. Patel, J. L. Cabellos, L. E. Rosas-Hernandez, K. Reynolds, H. O. H. Churchill, S. Barraza-Lopez, B. S. Mendoza, H. Nakamura, Substrate Interference and Strain in the Second-Harmonic Generation from MoSe2 Monolayers. *Nano Lett.*, doi: 10.1021/acs.nanolett.4c03880 (2024).

41. Y. Wu, Q. Tong, G.-B. Liu, H. Yu, W. Yao, Spin-valley qubit in nanostructures of monolayer semiconductors: Optical control and hyperfine interaction. *Phys. Rev. B* **93**, 045313 (2016).

42. Z. Gong, G.-B. Liu, H. Yu, D. Xiao, X. Cui, X. Xu, W. Yao, Magnetoelectric effects and valley-controlled spin quantum gates in transition metal dichalcogenide bilayers. *Nat. Commun.* **4**, 2053 (2013).

43. J. Pawłowski, J. E. Tiessen, R. Dax, J. Shi, Electrical manipulation of valley qubit and valley geometric phase in lateral monolayer heterostructures. *Phys. Rev. B* **109**, 045411 (2024).

44. F. Casola, T. Van Der Sar, A. Yacoby, Probing condensed matter physics with magnetometry based on nitrogen-vacancy centres in diamond. *Nat. Rev. Mater.* **3** (2018).

45. A. J. Healey, S. C. Scholten, T. Yang, J. A. Scott, G. J. Abrahams, I. O. Robertson, X. F. Hou, Y. F. Guo, S. Rahman, Y. Lu, M. Kianinia, I. Aharonovich, J.-P. Tetienne, Quantum microscopy with van der Waals heterostructures. *Nat. Phys.* **19**, 87–91 (2023).

**Acknowledgments:** We acknowledge discussions with A. MacDonald, C. Lei, N. Morales-Durán, P. Potasz, F. Wang and K. F. Mak.

**Funding:** This work was supported by the US Department of Energy (DOE), Office of Science, Basic Energy Sciences (BES), Materials Sciences and Engineering Division under award number DE-SC0020115 and, for sample preparation, under SLAC FWP 100459. Additional support was provided for sample preparation and analysis by the Betty and Gordon Moore Foundation EPiQS Initiative through Grant No. GBMF9462 and for a graduate fellowship for J. H. by NTT Research. Sample fabrication was performed at the Stanford Nano Shared Facilities (SNSF), supported by the National Science Foundation under award ECCS-2026822. K.W. and T.T. acknowledge support from the JSPS KAKENHI (Grant Numbers 21H05233 and 23H02052) and World Premier International Research Center Initiative (WPI), MEXT, Japan.

**Author contributions:** LY and TH conceived the project. KP and JH fabricated the samples. LY conducted the optical measurements with assistance from KP and JH. SK performed the moiré electronic structure calculations. LY developed the theoretical $\mathbf{k} \cdot \mathbf{p}$ framework to deduce spin relaxation from DFT-generated wavefunctions. TT and KW synthesized the h-BN crystals. FdJ and TH supervised the project. LY wrote the manuscript.

**Competing interests:** Authors declare that they have no competing interests.

**Data, code, and materials availability:** The datasets generated during and/or analyzed during this study are available from the corresponding author upon reasonable request.

**Supplementary Materials**

Materials and Methods

Supplementary Text

Figs. S1 to S17

References (*1–83*)

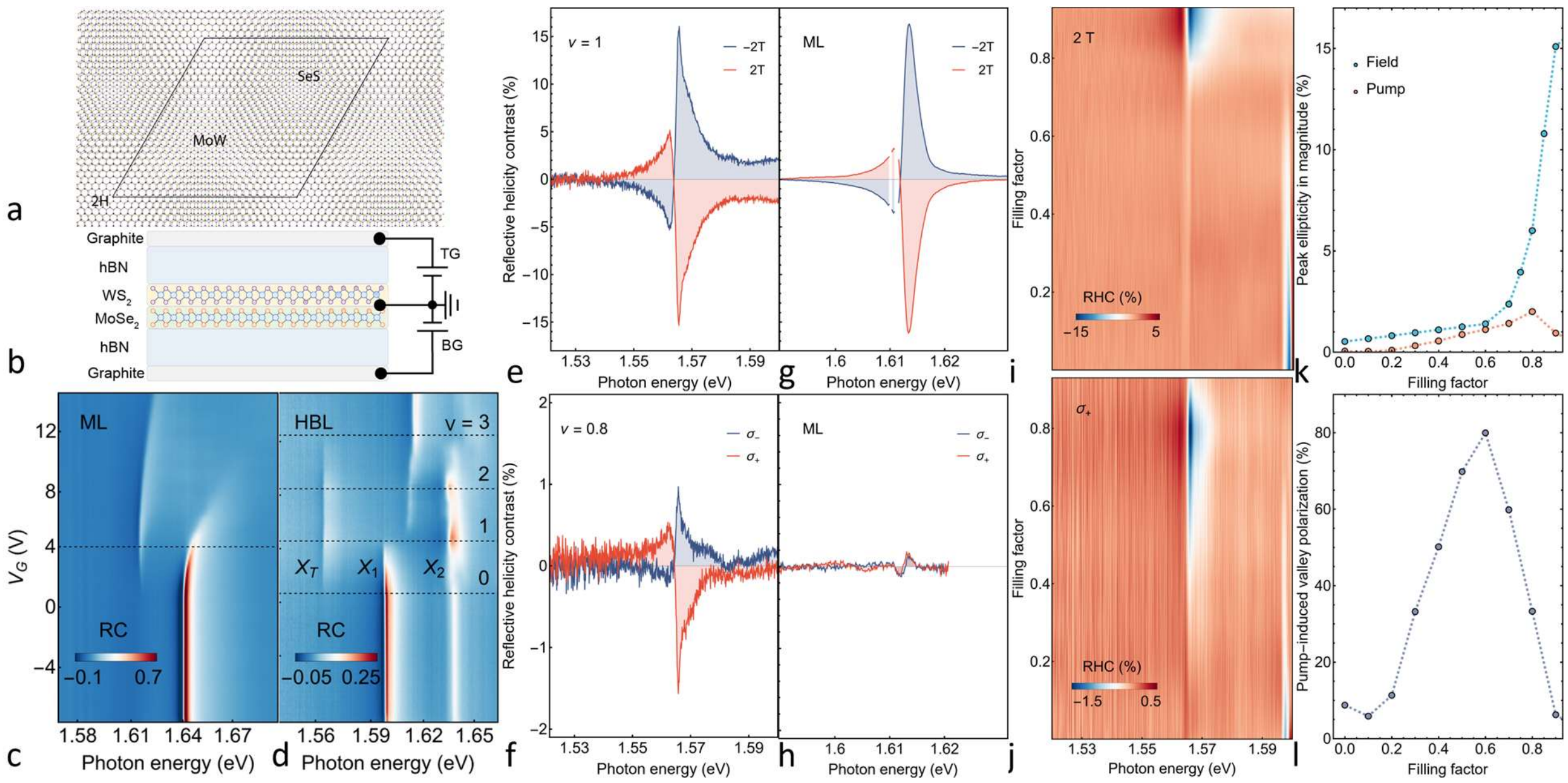

**Figure 1. a,** Moiré superlattice formed in angle-aligned $MoSe_2/WS_2$ heterobilayers with the superlattice cell outlined in black. The high-symmetry stackings are identified as 2H, MoW, and SeS, also referred to as AA', AB', and A'B, respectively. **b,** The dual-gated device structure of the HBL. **c,** Doping-dependent reflection contrast spectra of a $MoSe_2$ ML feature the exciton and the trion peak. The dashed line marks the doping level at which subsequent helicity-contrasting spectra were recorded. **d,** Doping-dependent reflection contrast spectra of the HBL, showing distinct spectral feature changes at integer fillings of moiré sites (dashed lines). **e, f,** Reflective helicity contrast (RHC) spectra from the HBL with valley polarization induced by a magnetic field (**e**) or a pump light (**f**). The c.w. pump light for **f** was tuned to 1.64 eV in photon energy and set above the saturation power (20 nW). **g, h,** Similar spectra obtained from the ML. The helicity-dependent change in **h** is negligible within experimental uncertainty, despite using a c.w. pump light of 150 μW power, which is nearly four orders of magnitude higher than that used for the HBL. **i, j,** Doping dependence of the RHC spectra from the HBL, with valley polarization induced by a magnetic field (**i**) or a pump light (**j**). **k,** Peak ellipticity extracted from **i, j**. **l,** Pump-induced valley polarization estimated based on the ratio of the peak RHC in **k**.

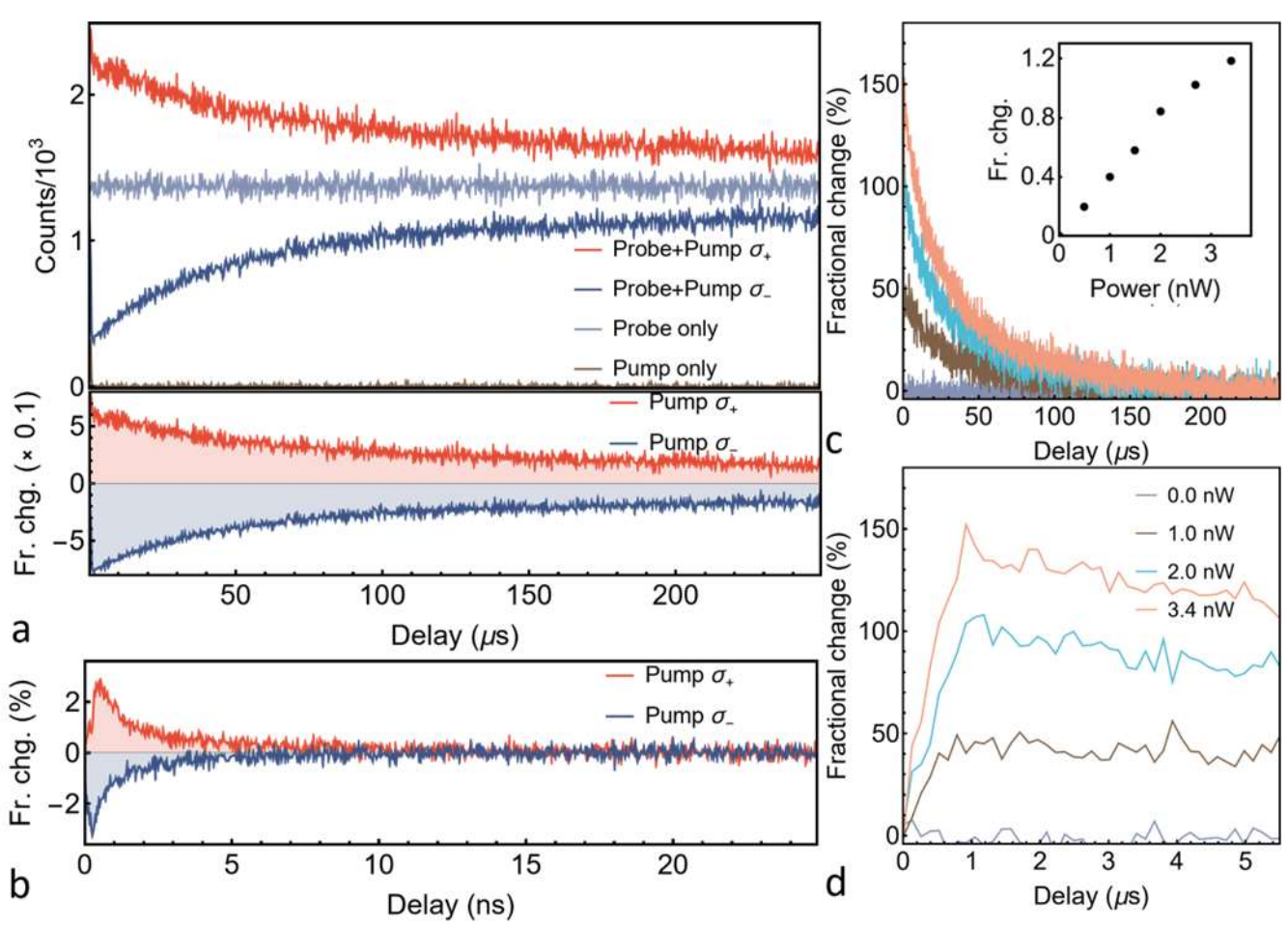

**Figure 2. Time-resolved light ellipticity in heterobilayer and monolayer $MoSe_2$. a,** Temporal profiles of probe-light counts from the heterobilayer (HBL) under various configurations of pump and probe light. In the pump-only trace, the peak near time zero is attributed to the PL from the trion resonance, lasting for the duration of the pump light (i.e., 1 μs). The photon energy of the pump light was 1.64 eV. The filling factor of the HBL was set at 0.8. Bottom panel: Fractional change in the probe light intensity, normalized to the level when the pump light is off. **b,** Fractional change in the probe light intensity following rapid spin relaxation in the ML, measured in nanoseconds. The broadband pump light was filtered to a photon energy range of 1.68-1.75 eV and set to an average power of 85 μW for the $\sigma_-$ configuration. The slight asymmetry between the two traces is attributed to the difference in the pump power used. **c,** Fractional change in the HBL measured at various (average) pump powers in unit of nW. Inset: The peak fractional change increases linearly with the pump power up to 2 nW, after which it exhibits sub-linear growth. **d,** Enlarged view of the data in **c** with reduced time spans, to resolve the gradual buildup of valley polarization over the 1-μs duration of the pump pulse.

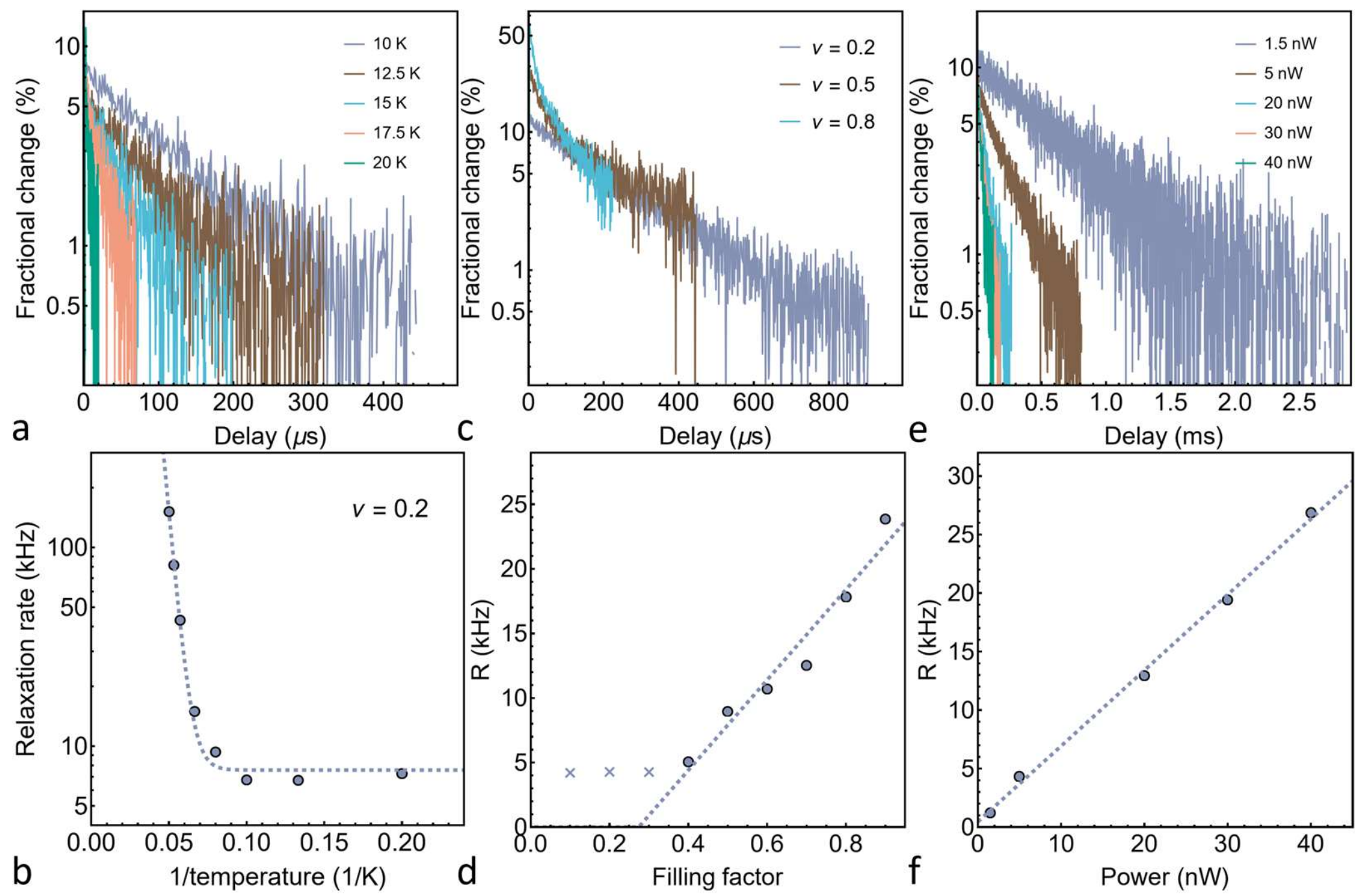

**Figure 3. a,** The fractional change in the probe light is shown as a function of the time delay after the pump pulse at five representative temperatures, with a representative filling factor of 0.2 in the low doping regime and with a probe power of 5 nW. **b,** Temperature dependence of the spin relaxation rate. The Arrhenius fit yields an activation energy of 17 meV. **c,** The same quantity as in **a** is shown for three representative filling factors, measured at a probe power of 5 nW. **d,** Doping dependence of the spin relaxation rate. The cross symbols indicate the rates obscured by additional photo-induced relaxation. **e,** The same quantity as in **a** is shown for various probe powers. The relaxation time at the probe power of 1.5 nW is 820 μs. **f,** Probe power dependence of the spin relaxation rate at a representative filling factor of 0.2, with the extrapolated relaxation rate at 0 nW being 0.46 kHz, corresponding to a relaxation time approaching 2 ms.

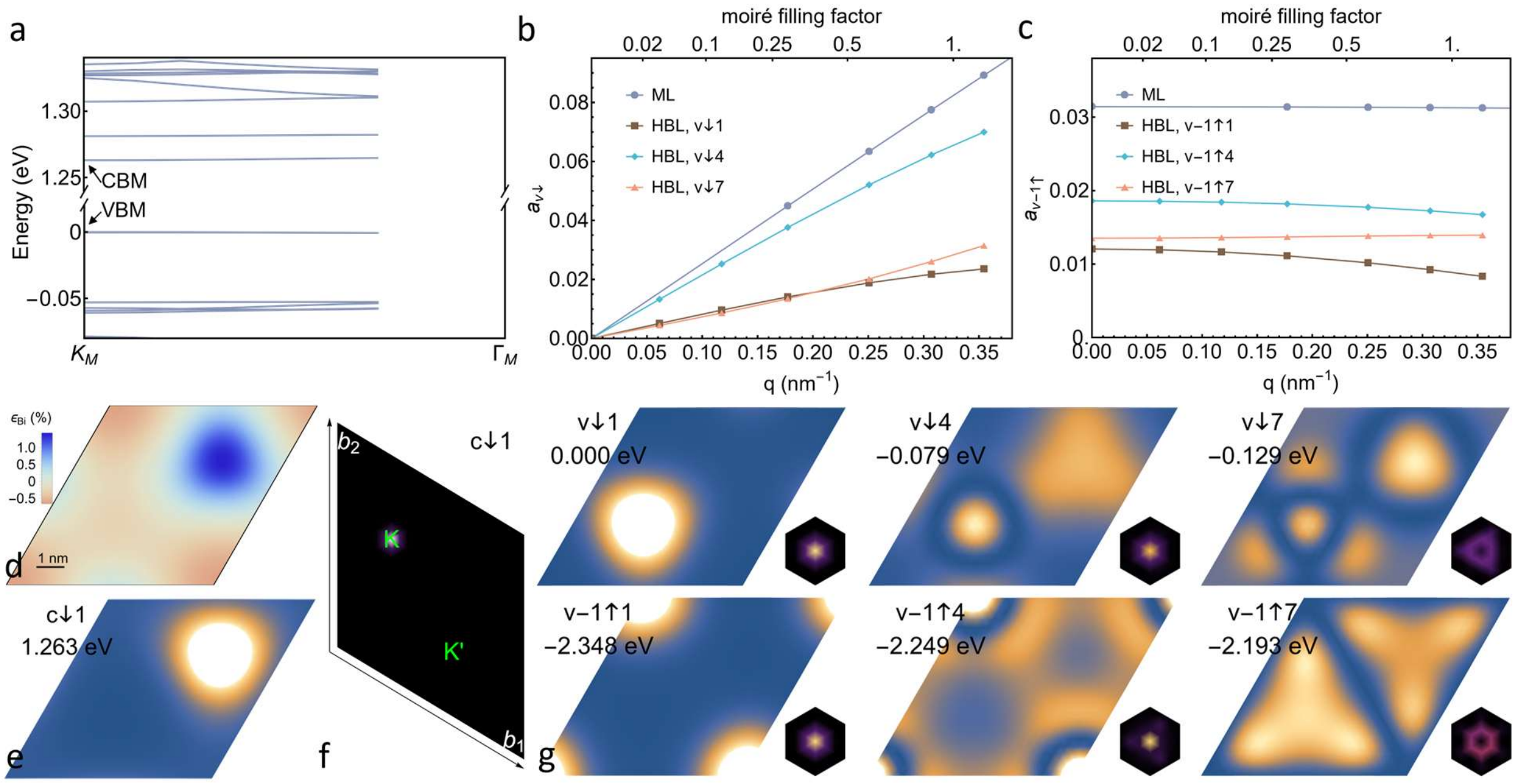

**Figure 4. a,** Band structure of the HBL near the CBM and VBM plotted in the moiré Brillouin zone. Band folding and the strain profile (shown in **d**) lead to energy-separated flat bands. The CBM is at $\mathbf{K}_\mathrm{M}$, where the effective mass roughly doubles. **b, c,** Projection amplitudes onto the $v\downarrow$ (**b**) and $v-1\uparrow$ (**c**) band for a wavefunction near CBM, as a function of carrier momentum along $\mathbf{K}_\mathrm{M}-\mathbf{\Gamma}_\mathrm{M}$ (bottom axis) or moiré filling factor (top axis). Compared to the ML result (gray line), the HBL results are labeled with the mini-band index (1, 4, 7) and lower in magnitude (see text). **d,** Biaxial strain distribution in the $MoSe_2$ part of the HBL over a moiré supercell from lattice reconstruction. Both tensile (blue) and compressive (red) strains are present, ensuring a net zero strain within the supercell. Scale bar: Strain values as a percentage change in the local lattice parameter from equilibrium. **e,** Spatial profile of the envelope function of the CBM wavefunction, maximizing in probability in the region of high tensile strain. **f,** Intensity map of the Fourier transform of the CBM wavefunction within one-third of the UBZ, showing a distinct peak at **K**. $\mathbf{b}_1$ and $\mathbf{b}_2$ denote the reciprocal lattice vectors. **g,** Envelope functions of the most significant KNN wavefunctions, which carry the highest projection amplitudes and contribute to spin relaxation over tenfold more than unlisted ones. The overlaps between many of these envelope functions are notably less than unity. Also annotated is the eigenenergy of the wavefunction relative to the VBM energy. Insets: Similar results as in **f** but zoomed in to the region up to the second nearest neighbors of **K**, where the neighbors are separated by primitive reciprocal moiré lattice vectors ($\mathbf{b}_{\mathrm{M},1}$ or $\mathbf{b}_{\mathrm{M},2}$).